# Probing blood pressure and arterial stiffness noninvasively by guided axial waves

Guo-Yang Li [*, †], Yuxuan Jiang, Yang Zheng, Weiqiang Xu, Zhaoyi Zhang, Yanping Cao[†]

Institute of Biomechanics and Medical Engineering, AML, Department of Engineering Mechanics, Tsinghua University, Beijing 100084, PR China

[*]Present address: Harvard Medical School and Wellman Center for Photomedicine, Massachusetts General Hospital, Boston, MA 02117, USA

[†]Corresponding author: G.Y. Li (lgy14@tsinghua.org.cn); Y. Cao (caoyanping@tsinghua.edu.cn)

*Abstract*: The clinical and economic burden of cardiovascular diseases (CVDs) poses a global challenge. Growing evidence suggests an early assessment of arterial stiffness can provide insights into the pathogenesis of CVDs. However, it remains difficult to quantitatively characterize the arterial stiffness *in vivo*. Here we utilize the guided elastic waves continuously excited and detected by ultrasound as a tool to probe the blood pressure (BP) and mechanical properties of the common carotid artery (CCA) simultaneously. In a pilot study of 17 healthy volunteers, we obverse a $\sim 20\%$ variation in the group velocity ($5.16 \pm 0.55$ m/s in systole and $4.31 \pm 0.49$ m/s in diastole) induced by variation of the BP. A linear relationship between the square of the group velocity and the BP is revealed by our experimental data and finite element analysis, which enables us to measure the waveform of the BP. Furthermore, we propose to use the wavelet analysis to extract the dispersion relation of the guided waves, which provides a quantitative measurement of the arterial stiffness. The results and methods reported in this study show the group velocity and dispersion relation of the guided waves can be adopted to probe BP and arterial stiffness noninvasively and thus promising for early diagnosis of the CVDs.

## 1  Introduction

Cardiovascular diseases (CVDs) are the number one cause of human death globally (Lindsay and Dietz, 2011; Pagidipati and Gaziano, 2013). People suffering from CVDs face not only this death threat but also the financial burdens of medications (Pandey and Meltzer, 2016). While blood pressure (BP) may be the most common metrics for assessment of cardiovascular health (Dzau and Balatbat, 2019; Ma et al., 2018; Psaty et al., 2001; Wang et al., 2018), more and more emphasis has also been placed on the role of arterial stiffness in the development of CVDs (Laurent et al., 2006; Safar, 2018; Vasan et al., 2019; Zieman et al., 2005). In the pathogenesis of CVDs, the interplay between BP and arterial stiffness is actually complex; arteries remodel in hypertension to adapt to the increased BP; the increased arterial stiffness, in turn, contributes to the development and complications of hypertension (Humphrey, 2008; Intengan and Schiffrin, 2001; Mitchell, 2014). Monitoring the BP and arterial stiffness independently in a noninvasive manner thus holds promises for gaining new insights into the pathogenesis of CVDs and the early prediction of cardiovascular disease events but remains a challenging issue to date.

The beating heart creates blood pressure and flow pulsations that propagate as waves through the arterial tree (Vosse and Stergiopulos, 2011). The pulse wave velocity (PWV), which is thus determined by not only the mechanical properties of the arterial wall but also the dynamics of the blood flow (Mitchell, 2014; Vosse and Stergiopulos, 2011), has been adopted as a measure of the arterial stiffness (Chirinos, 2012; Laurent et al., 2006; Spronck and Humphrey, 2019; Vappou et al., 2010). Although the PWV measurements are now easy to be implemented by measuring the time delays of pulse waveforms along a short arterial segment (Luo et al., 2012; Mirault et al., 2015) or tree (Chirinos, 2012), they are found only moderately correlated because other factors (e.g., dynamics of the blood flow, BP, and arterial geometries) that may come into play have not been fully taken into consideration, making the PWV play a complementary role in risk prediction (Mitchell, 2014).

To address the limitations of the PWV and quantitatively characterize the mechanical properties of the arteries, ultrasound elastography based on the guided waves excitation and detection in arterial walls has been proposed (Bernal et al., 2011; Couade et al., 2010; Li et al., 2017c). In contrast with the pulse wave, the frequencies of the guided waves excited remotely by acoustic radiation force (ARF) (Bercoff et al., 2004; Sarvazyan et al., 1998) are hundreds of times higher. Therefore, multi excitations and detections within one cardiac cycle are possible (Cikes et al., 2014; Couade et al., 2011), which gives access to interrogate the variations of the arterial stiffness induced by the BP *in vivo*. Due to the nonlinear mechanical properties of the arteries, which have been widely studied by the conventional mechanical characterization techniques (e.g., tensile (Fung, 1967) and inflation test (Hayashi et al., 1980)) and constitutive modeling (Gasser et al., 2006; Holzapfel et al., 2000), the instantaneous elastic properties of the arterial wall and thus the propagation speed of the guided waves, are expected to change with the BP (Couade et al., 2010; Li et al., 2017b). In this case, an appropriate wave motion model that incorporates the effect of the wall stress is a key to interpret the experimental results and infer the BP and arterial stiffness simultaneously.

In this work, we utilize the guided waves as a tool to probe the BP and mechanical properties of the common carotid artery (CCA), which is the easiest accessible artery *in vivo* by ultrasound imaging. We propose an imaging sequence that enables us to continuously excite and measure the guided waves in the CCA. A pilot study of 17 healthy subjects reveals a simple correlation between the group velocity and the BP, which is confirmed by the finite element analysis (FEA). With the coefficients calibrated by the blood pressure monitor, the variation of the group velocity is interpreted as the waveform of the BP. We further propose a wavelet analysis-based method to robustly extract the dispersion relations of the guided waves obtained at different BPs, which enables us to quantitatively characterize the mechanical properties of the CCA.

## 2 Methods

### 2.1 Study design

We recruited 17 healthy volunteers (10 male, 7 female, age ranges from 22 to 28). Informed consent was obtained from each subject after an explanation of the nature and possible consequences of the study. Before and after the experiment, we measured the blood pressure and heart rate of each volunteer with a blood pressure monitor (Omron 5 Series Upper Arm Blood Pressure Monitor, Model BP7200). To keep the body position, we asked the volunteers to sit upright and look straight ahead during the experiment. The ultrasound transducer was placed along the axial direction of the right CCA (see Fig. 1A). In this orientation, the anterior and posterior walls that are parallel to each other can be observed in the grayscale image (see Fig. 1B). We relied on the ARF to remotely excite guided elastic waves within the arterial wall and then the ultrafast ultrasound imaging technique to monitor the wave propagation (see the imaging protocol for details). We limited the ultrasound intensity to ensure compliance with the requirements of the Food and Drug Administration (FDA), as described in Ref. (Li et al., 2019).

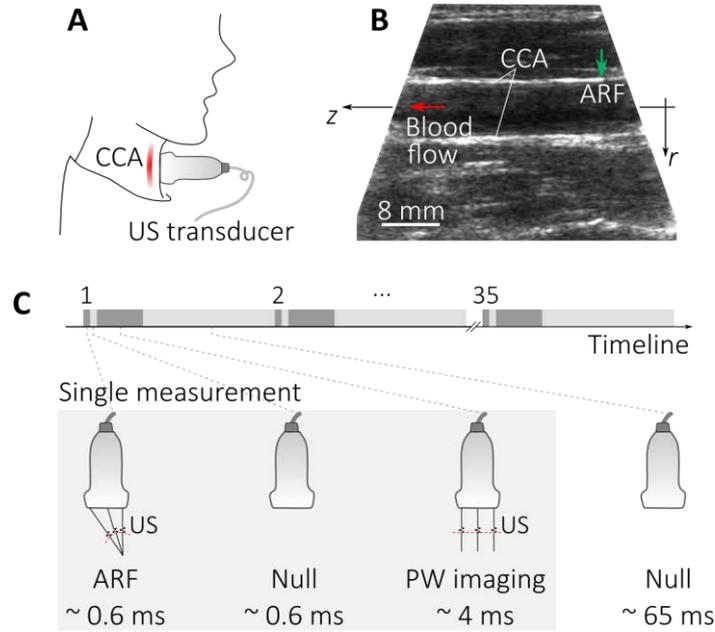

**Fig. 1 Excitation and measurement of the guided waves in common carotid artery.** (**A**) Schematic of the experiment. (**B**) Grayscale image of the CCA showing the location of ARF and direction of the blood flow. ARF is applied to the anterior wall. (**C**) Imaging sequence used in this study. Frame rate of the PW imaging is 10,000 Hz.

The protocol was approved by the institutional review board at Tsinghua University.

## 2.2 Imaging protocol

The experimental system was built on a Verasonics Vantage 64LE System (Verasonics Inc., Kirkland, WA, USA), equipped with a L9-4 transducer (JiaRui Electronics Technology Co., Ltd., Shenzhen, China). The central frequency and element number of the transducer were 6.5 MHz and 128, respectively. Custom beam sequences shown in Fig. 1C were implemented to perform the imaging. The ARF was generated by a focused ultrasound beam using 64 elements. The focal point of the beam moved vertically near the anterior wall (from ~3 mm above to ~3 mm below the anterior wall) in a very short time to induce elastic Cherenkov effect (Bercoff et al., 2004) and thus excite quasi-plane shear wave. The duration at each location was ~50 $\mu$s (300 cycles). The total duration of the excitation process was ~0.6 ms. About ~0.6 ms after the excitation, the transducer started to acquire the in-phase and quadrature (IQ) data at a frame rate of 10,000 Hz. We used one transmit/receive event for each frame and acquired in total 40 frames (~4 ms). All the 128 elements were used to transmit but only 64 elements at the central part of the transducer were used to receive. The plane wave imaging with delay and sum beamforming was adopted to reconstruct each frame (Montaldo et al., 2009). Based on the IQ data the particle velocity field $v_r(z,t)$ was calculated offline with the Loupas' estimator (Loupas et al., 1995). Here $r$ and $z$ denote the cylindrical coordinate system showing in Fig. 1B, and $t$ denotes the time. This single measurement (guided wave excitation and tracking) takes $\sim 5$ ms, which is far shorter than the time of one cardiac cycle ($\sim 800$ ms). We thus can suppose the deformation induced by pulse wave is quasi-static and the guided waves excited by ARF is an incremental deformation on the static deformation. To continuously measure the mechanical property of the arterial wall in a cardiac cycle, we repeated the measurement for $35$ times (see Fig. 1C). The time between each measurement was $\sim 65$ ms. In total the $35$ measurements take $\sim 2.45$ s, covering approximately $3$ cardiac cycles.

## 2.3 Group velocity measurement

We performed the Radon transformation of $v_r(z,t)$ to estimate the group velocity. The Radon transformation is defined as

$$\hat{v}_r(\varphi, \bar{\rho}) = \int_{l:z\cos\varphi - t\sin\varphi = \bar{\rho}} v_r(z,t) \mathrm{d}l, \qquad (1)$$

where $\tan\varphi$ and $\bar{\rho}$ denote the slope and intercept of the line $z\cos\varphi - t\sin\varphi = \bar{\rho}$, along which the values of $v_r(z,t)$ are summed. When the line is aligned with the wavefront in $z-t$ plane, $\hat{v}_r(\varphi, \bar{\rho})$ reaches the maximum (see supplementary Fig. S1) and then the group velocity can be calculated by

$$c_g = \tan\varphi. \qquad (2)$$

## 2.4 Finite element analysis

The BP induces the variation of the circumferential stress that changes the stiffness of the arterial wall and thus the group velocity. We carried out FEA to study the effects of BP on the group velocity using the commercial FEA software Abaqus 6.13 (Dassault Systems, Waltham, MA, USA). The material behavior of the arterial wall is modeled by the Holzapfel-Gasser-Ogden (HGO) model (Gasser et al., 2006), for which the strain energy function is

$$W = C_{10}(I_1 - 3) + \frac{k_1}{2k_2} \sum_{i=1}^{2} \{e^{k_2[\kappa(I_1-3)+(1-3\kappa)(I_{4i}-1)]^2} - 1\}, \tag{3}$$

where

$$I_1 = \text{tr}(\mathbf{F}^T\mathbf{F}), I_{41} = \text{tr}(\mathbf{M}_1 \cdot \mathbf{F}^T\mathbf{F}\mathbf{M}_1), I_{42} = \text{tr}(\mathbf{M}_2 \cdot \mathbf{F}^T\mathbf{F}\mathbf{M}_2). \tag{4}$$

$\mathbf{F}$ is the deformation gradient. The two unit-vectors $\mathbf{M}_1$ and $\mathbf{M}_2$ denote the mean orientations of the two families of collagen fibers (see supplementary Fig. S2). The values of the constitutive parameters $C_{10}$, $k_1$, $k_2$, and $\kappa$ were taken from Ref. (Gasser et al., 2006): $C_{10} = 7.64$ kPa, $k_1 = 996.6$ kPa, $\kappa = 0.226$, and $k_2 = 524.6$ (from 200 to 600). The geometry curvature that only came into play in the low frequency regime(Li et al., 2017a) was ignored for simplification. The thickness and in-plane size of the model were 1 mm and 25 mm × 40 mm. We applied a tensile stress $\sigma_{\theta\theta}$ along the circumferential direction and restricted the axial displacement to model the stress state induced by the BP. The tensile stress $\sigma_{\theta\theta}$ in the arterial wall estimated by

$$\sigma_{\theta\theta} = \text{BP} \cdot R/h, \tag{5}$$

is usually no more than 60 kPa, where $R$ and $h$ denote the inner radius and wall thickness of the artery, respectively. We thus varied the dimensionless parameter $\sigma_{\theta\theta}/2C_{10}$ from 0 to 4 in the FEA. A constant body force (duration 0.2 ms) with a Gaussian distribution was used to model the ARF induced by an ultrasound beam (diameter 1.0 mm). In total 234,000 C3D8R elements were used in the simulations. Computation convergence was confirmed by refining the mesh and time increment.

## 2.5 Wavelet analysis to extract the dispersion relation

The ARF can excite guided waves in the arterial wall over a frequency range centered around 1 kHz. In this frequency range, it is the lowest order quasi-asymmetric mode (denoted by $A_0$) that is predominantly excited (Li et al., 2017a). In previous studies, two-dimensional Fourier transformation of the particle velocity field $v_r(z,t)$ was widely used to extract the dispersion relation of $A_0$ (Bernal et al., 2011; Couade et al., 2010; Li et al., 2017a). However, this algorithm becomes less robust when pulse wave motion comes into play. Here we propose to extract the dispersion relation using wavelet analysis of the particle velocities. Only the particle velocities of two points $v_r^a(t)$ and $v_r^b(t)$, where $a$ and $b$ denote two positions in the anterior wall, separated by a distance $d_{ab}$ in the axial direction, are needed to perform this analysis. The continuous wavelet transformations of $v_r^a(t)$ and $v_r^b(t)$ are

$$V_r^a(\tau, \alpha) = \frac{1}{\sqrt{\alpha}} \int_{-\infty}^{+\infty} v_r^a(t) \, \phi^*\left(\frac{t-\tau}{\alpha}\right) dt,$$

$$V_r^b(\tau, \alpha) = \frac{1}{\sqrt{\alpha}} \int_{-\infty}^{+\infty} v_r^b(t) \, \phi^*\left(\frac{t-\tau}{\alpha}\right) dt,$$
(6)

where $\tau$ and $\alpha$ denote the time delay and a scale, respectively. $\phi$ is the Morlet wavelet (Kijanka et al., 2019; Wu et al., 2009) and $\phi^*$ denotes the complex conjugate of $\phi$,

$$\phi(t) = \exp\left(-\frac{t^2}{2\sigma^2}\right) \exp(i 2\pi f_c t), \qquad (7)$$

where $\sigma = \sqrt{3}$ and $f_c = 1$ are two parameters of the Morlet wavelet.

For a given scale $\alpha$, we calculated the cross-correlation of $V_r^a$ and $V_r^b$; the phase of the cross-correlation coefficient reaches a minimum when $V_r^a$ and $V_r^b$ are in phase. In this way we can obtain the phase difference (or the time delay $\Delta\tau$) between $a$ and $b$. The phase velocity $c$ at this frequency ($f_c/\alpha$) is thus be computed by $c = d_{ab}/\Delta\tau$. By varying the scale $\alpha$ we can extract the dispersion relation of $A_0$. We validated the proposed algorithm based on the data produced by FEA (see Supplementary Fig. S3). The distance $d_{ab}$ is recommended to be comparable to the wavelength at the central frequency. In the present study $d_{ab} = 5$ mm.

### 2.6 Statistical analysis

Data fitting and statistical analysis were performed using Matlab R2017b (MathWorks Inc., Natick, MA, USA). The $r^2$ (coefficient of determination) was used as a measure of the linear/nonlinear least squares regressions. Student's *t*-test was performed to compare the shear modulus of male and female volunteers. A $p$ value of 0.05 was adopted to indicate statistical significance.

## 3 Results

### 3.1 Guided waves measured in the common carotid artery (CCA)

Figure 2A is a typical map of the radial particle velocity $v_r(z, t)$, showing the wave propagation along the center of the anterior wall. The solid curves overlaid on this map are the time profiles of the particle velocity $v_r$ at two points (denoted by $v_r^a(t)$ and $v_r^b(t)$, respectively) separated by $d_{ab} \approx 5$ mm. We note the shape variation between the time profiles $v_r^a(t)$ and $v_r^b(t)$, which suggests the wave dispersion is apparent.

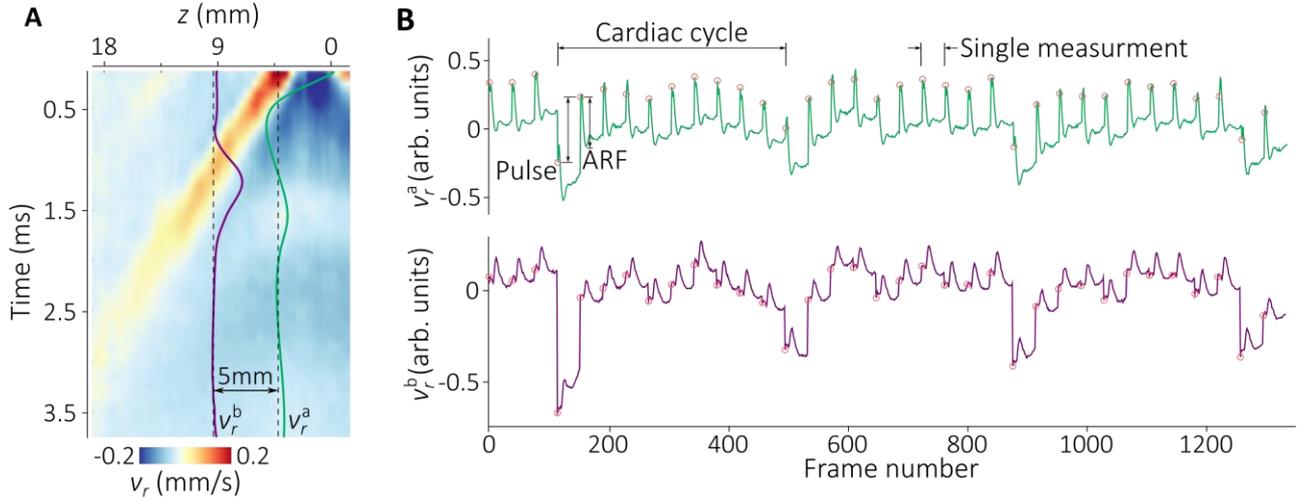

**Fig. 2 Guided waves measured in an experiment.** (**A**) Time-space domain data obtained from a single measurement in diastole showing the guided axial wave propagation in the anterior wall. The two solid lines are time profiles of the radial particle velocity ($v_r^a$ and $v_r^b$) at two points in the anterior wall. The distance between the two points $d_{ab} \approx 5$ mm. (**B**) Time profiles of $v_r^a$ and $v_r^b$ obtained from the 35 continuous measurements. Note the time gap (~65 ms) between each measure is not shown in this plot. The wave amplitudes of $v_r$ induced by ARF and the pulse wave respectively are comparable to each other, as shown on the curve of $v_r^a$. For this subject, the HR is ~85 beats/min. Therefore ~10 measurements were performed during each cardiac cycle.

For the subject shown in Fig. 2A, the time profiles of $v_r^a$ and $v_r^b$ obtained by the 35 continuous measurements are shown in Fig. 2B. It should be noted that the time gaps (~65 ms) between adjacent measurements are not shown in this figure. The periodic variations of the time profiles are synchronous with the cardiac cycles. In each cardiac cycle, the shifts of the time profiles were induced by motions of the arterial walls. Comparing the two curves we note the shifts of the profiles $v_r^a(t)$ and $v_r^b(t)$ were approximately the same, suggesting the pulse wave approximately induces a constant particle velocity in the anterior wall. This can be understood by the fact that the wavelength of the pulse wave is significantly longer than $d_{ab}$ (Vappou et al., 2010). In other word, the BP during a single measurement can be supposed as a constant and different measurements in a cardiac cycle are performed under different BP levels. While the deformation induced by the ARF (the displacement is on the order of μm) is much smaller than that induced by the pulse wave (on the order of mm), we note the wave amplitude in $v_r$ induced by pulse wave is comparable to that induced by ARF (see Fig. 2B). This makes the guided waves induced by the ARF be detectable at different time in a cardiac cycle. The amplitude in $v_r$ induced by the pulse wave was removed by a digital high pass filter in our post-processing.

### 3.2 Variation of the group velocity in a cardiac cycle

The arterial wall is a nonlinear material with its stiffness increasing with strain (Fung et al., 1979; Wagenseil and Mecham, 2009), hence, we can expect the BP variation may have effects on the guided wave motion. To test this hypothesis, we studied the variations of the group velocity for each subject. Typical results for three subjects (#1, #2, and #3) are shown in Fig. 3A. For each subject we also report the heart rate (HR), systolic BP (SBP), and diastolic BP (DBP), which were acquired by a BP monitor (see Methods). Interestingly, we find the group velocity of the guided axial wave (denoted by $c_g$) periodically varies in cardiac cycles. With the help of the grayscale maps, we also measured the variations of the arterial diameter $2R$, which were induced by the pulse waves (see Supplementary Fig. S4). The variation of $c_g$ is in synchronization with the arterial diameter $2R$, with an apparent positive correlation. This experimental observation agrees well with the hypothesis. Higher BP enlarges the arterial stiffness and diameter; the increased arterial stiffness thus leads to the increase in $c_g$.

From the curves of the group velocities, we identified three cardiac cycles for each subject. For each cardiac cycle we found the maximum and minimum of $c_g$. Then we calculated the mean values of the three maximums and minimums (denoted by $c_g^{\max}$ and $c_g^{\min}$, respectively). As shown in Fig. 3B, the mean values of $c_g^{\max}$ and $c_g^{\min}$ for all the 17 subjects are $5.16 \pm 0.55$ m/s and $4.31 \pm 0.49$ m/s, suggesting a ~20% variation in $c_g$ can be detected for humans with normal BPs.

The FEA helps to quantitatively understand the effect of BP on $c_g$. Based on the particle velocity field obtained from the FEA, we calculated the group velocity $c_g$ using the Radon transformation algorithm. As shown in Fig. 3C, we report the square of group velocity $\rho c_g^2/2C_{10}$ as a function of the dimensionless circumferential stress $\sigma_{\theta\theta}/2C_{10}$. The FEA results clearly reveal a linear correlation between $\rho c_g^2/2C_{10}$ and $\sigma_{\theta\theta}/2C_{10}$ (the coefficient of determination $r^2 \sim 0.99$) when $\sigma_{\theta\theta}/2C_{10} < 4$, which suggests $\mathrm{BP} = \alpha c_g^2 + \beta$, where $\alpha$ and $\beta$ are the coefficients. We also note the coefficients $\alpha$ and $\beta$ are insensitive to the parameter $k_2$, which describes the nonlinear (exponential) stiffening effect (Gasser et al., 2006).

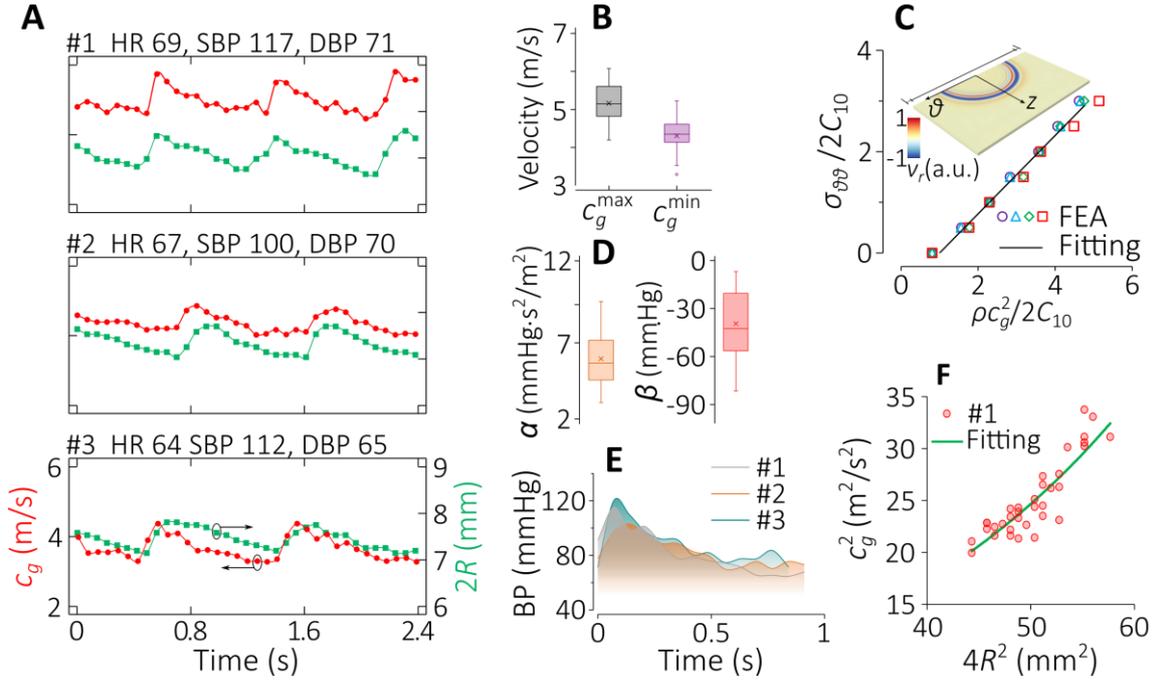

**Fig. 3 Variation of the group velocity in a cardiac cycle.** (**A**) The group velocities $c_g$ and arterial diameters $2R$ for three typical subjects (# 1, #2 and #3). For each subject, the heart rate (HR), systolic blood pressure (SBP) and diastolic blood pressure (DBP) are reported. (**B**) The statical results ($N = 17$) for group velocities. $c_g^{max}$ and $c_g^{min}$ denote the maximum and minimum of the group velocity for each subject. (**C**) Finite element analysis to illustrate the effect of blood pressure on the group velocity. Circle, $k_2 = 200$; Triangle, $k_2 = 300$; Diamond, $k_2 = 524.6$; Square, $k_2 = 600$. The results reveal a linear relation between the circumferential stress and the square of the group velocity. The slopes are insensitive to the parameter $k_2$. Solid line is the linear fitting for $k_2 = 524.6$: $\sigma_{\theta\theta}/2C_{10} = 0.773 \, \rho c_g^2/2C_{10} + 0.772$ ($r^2 \sim 0.99$). (**D**) The statical results ($N = 17$) for the coefficients $\alpha$ and $\beta$. $\mathrm{BP} = \alpha c_g^2 + \beta$. (**E**) The BP waveforms for the three subjects shown in (A). (**F**) Plot showing $c_g^2$ as a function of $4R^2$ for subject #1. The solid line is the best fitting with an exponential function ($r^2 \sim 0.79$).

The linear correlation confirmed by FEA provides us a convenient way to probe the BP waveform. Once the two coefficients $\alpha$ and $\beta$ are calibrated, we can directly interpret the group velocity as BP. For all the subjects, we calibrated the coefficients by

$$\alpha = (\text{SBP} - \text{DBP})/\left[\left(c_g^{\max}\right)^2 - \left(c_g^{\min}\right)^2\right], \tag{8}$$

$$\beta = \left[\left(c_g^{\max}\right)^2 \text{DBP} - \left(c_g^{\min}\right)^2 \text{SBP}\right]/\left[\left(c_g^{\max}\right)^2 - \left(c_g^{\min}\right)^2\right]. \tag{9}$$

Fig. 3D shows the statical results obtained from all the subjects. The mean values for $\alpha$ and $\beta$ obtained from the experiments are $5.8 \pm 1.7$ mmHg·s²/m² and $-39.7 \pm 23.0$ mmHg. For the three subjects shown in Fig. 3A, we calibrate the coefficients $\alpha$ and $\beta$ and plot the BP waveforms in Fig. 3E. Interestingly, we find the waveforms measured with our method show similar shapes as those reported in the literature (Takazawa et al., 2007; Wang et al., 2018).

Regarding the noninvasive measurement of BP waveform, the variation of the arterial diameter $2R$ induced by the pulse wave is a popular quantity to be used (Arndt et al., 1968; Wang et al., 2018). The theoretical foundation of these methods is an empirical formula that the BP is an exponent function of the arterial cross section. In Fig. 3F we also plot $c_g^2$ as a function of $4R^2$ for subject #1, which approximately is an exponential function ($r^2 \sim 0.79$) and thus further validates the linear relation between $c_g^2$ and BP.

### 3.3 Wavelet analysis to extract the dispersion relation of the guided wave

We proceed to analyze the dispersion relations of the guided waves. Quantitatively the arterial stiffness can be computed by fitting the dispersion relation of the $A_0$ mode with the Lamb wave model (Couade et al., 2010; Li et al., 2017b). However, in contrast to the Lamb wave model, the curvature of the arterial wall may have a pronounced effect on the dispersion relation in the low frequency regime (Couade et al., 2010; Li et al., 2017a). For guided axial waves, we have previously reported a critical frequency $f_c = 0.304 h^{-1}(E/3\rho)^{1/2}(R/h)^{-0.934}$, beyond which the effect of the curvature is negligible (Li et al., 2017a). For CCA with $R \sim 4$ mm, the wall thickness $h \sim 0.9$ mm, and the elastic modulus $E \sim 200$ kPa, the critical frequency is approximately $0.6$ kHz. Therefore, only the dispersion relation in the frequency range of $0.6$ kHz to $1.6$ kHz is adopted for subsequent analysis.

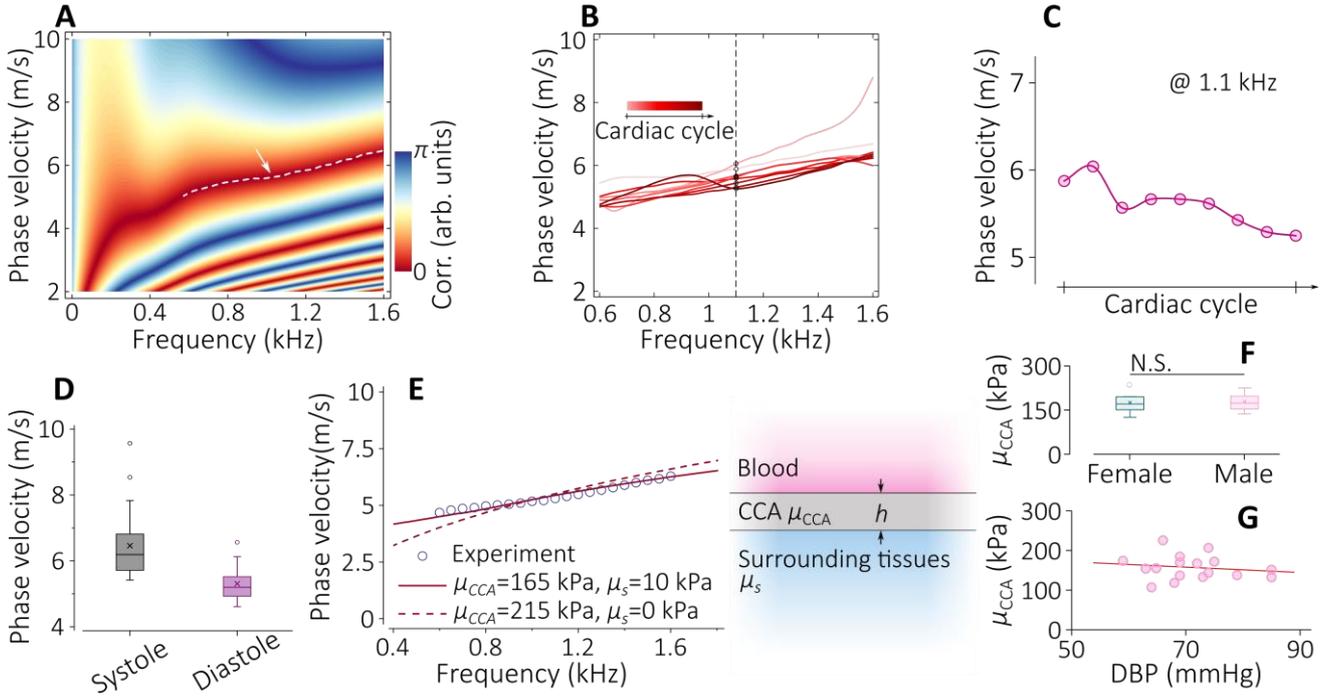

**Fig. 4 Dispersion relations extracted by wavelet analysis.** (**A**) Cross-correlogram of the wavelet signals. The dispersion relation of the $A_0$ mode, which is highlighted by the dashed white line, can be identified by finding the minimums in the map. The data in the low frequency regime (< 600 Hz) is abandoned. (**B**) Dispersion relations obtained in a cardiac cycle. The HR for this subject is 90. (**C**) Phase velocities at 1.1 kHz extracted from (B). (**D**) Statistical analysis of the phase velocities (@ 1.1 kHz) in systole and diastole for all the 17 subjects. (**E**) Fitting of a typical dispersion relation obtained in diastole with the three-layer model, in which the CCA and the surrounding tissues are assumed to be isotropic, incompressible, and stress-free materials. A better fitting is achieved when the stiffness of the surrounding tissues ($\mu_s$) is taken into consideration. $h = 0.95$ mm. (**F**) The shear modulus of the CCA ($\mu_{CCA}$) obtained in diastole for female and male volunteers. No statistically significant difference is found between the male and female volunteers ($p > 0.05$). (**G**) Correlation analysis between $\mu_{CCA}$ and DBP. No correlation between $\mu_{CCA}$ and DBP is found (correlation coefficient $-0.15$). Points: experimental data, Solid line: linear regression of the experimental data.

The dispersion relation of the $A_0$ mode was extracted by the wavelet analysis method (see Methods for details). Figure 4A shows a typical cross-correlogram (i.e., the phase of the cross-correlation of $V_r^a(\tau,\alpha)$ and $V_r^b(\tau,\alpha)$). At each frequency we can find multi minimums; the one that corresponds to the $A_0$ mode can easily be identified by searching in the range of 4 to 10 m/s. In this way we can extract the dispersion relation, as depicted by the dashed line. Figure 4B shows the dispersion relations obtained in one cardiac cycle, from which we find the BD has a pronounced effect on the phase velocity. The variations of the phase velocity in one cardiac cycle at 1.1 kHz is shown in Fig. 4C. Similar to the group velocity, the phase velocity reaches the maximum/minimum in systole/diastole. For all the 17 subjects, the phase velocities were $6.48 \pm 1.06$ m/s in systole and $5.30 \pm 0.57$ m/s diastole (mean ± SD), as shown in Fig. 4D. Again a ~20% variation of the phase velocity induced by the BP was observed. Comparing with the group velocity, the phase velocity at 1.1 kHz is larger because the central frequency of the guided waves excited by the ARF is lower than 1.1 kHz. The two-dimensional Fourier transformation of the time-space domain data shows the wave energy is dominated by these wave modes with frequencies lower than 0.8 kHz (see Supplementary Fig. S5), indicting the group velocity is primarily determined by these low frequency waves that have a lower phase velocity.

The dispersion relations enable us to quantitively interrogate the mechanical properties of the arterial walls. For the sake of simplicity, we adopted the leaky Lamb wave model that incorporates the effects of the blood and the elasticity of the surrounding tissues (Li et al., 2019) to fit the dispersion relations obtained in diastole. As shown in Fig. 4E, a better fitting of the experimental data was achieved when both the elasticities of the arterial wall ($\mu_{CCA}$) and the surrounding tissues ($\mu_s$) were taken into consideration. Modeling the surrounding tissues as fluid ($\mu_s$ is zero) results in a significant overestimation of the arterial stiffness. Although $\mu_s$ should be slightly different for each subject, a constant value $\mu_s = 10$ kPa estimated from the shear wave velocity of the surrounding tissue, was taken to make $\mu_{CCA}$ was the only fitting parameter (see Supplementary Fig. S6 and Note1). Statistical analysis was then performed on all the 17 healthy volunteers. As shown in Fig. 4F, no significant difference between the shear modulus $\mu_{CCA}$ of male ($168.0 \pm 30.4$ kPa) and female ($164.3 \pm 40.3$ kPa) is found. The correlation analysis between the shear modulus and DBP yields a correlation coefficient $-0.15$ (Fig. 4F), which suggests the shear modulus shows no dependence on DBP.

## 4    Discussion

Monitoring the BP and arterial stiffness quantitatively holds promises for early prediction of cardiovascular disease events. In this study, we have shown that imaging the guided axial waves excited remotely by ultrasound provides a unique way for probing the BP and arterial stiffness *in vivo*. Multi measurements (~10 times) within one cardiac cycle gives us the opportunity to study the variation of the group velocity and dispersion relation. The group velocities of the CCA in systole and diastole measured from healthy volunteers were $5.16 \pm 0.55$ m/s and $4.31 \pm 0.49$ m/s, respectively. The ~20% variation in the group velocity, together with the linear relationship between the square of the group velocity and the BP, provides a novel method to probe the local BP of the large arteries. The SBP and DBP we used to calibrate the coefficients $\alpha$ and $\beta$ in the linear relationship were obtained from the upper arm by a blood pressure monitor. While the DBP at CCA and brachial artery can be assumed the same, the brachial SBP must be calibrated before being interpreted as the SBP at CCA (Takazawa et al., 2007; Verbeke et al., 2005). With the waveform of the local BP measured by the group velocity, such a calibration is possible by the method proposed by Kelly and Fitchett (Kelly and Fitchett, 1992).

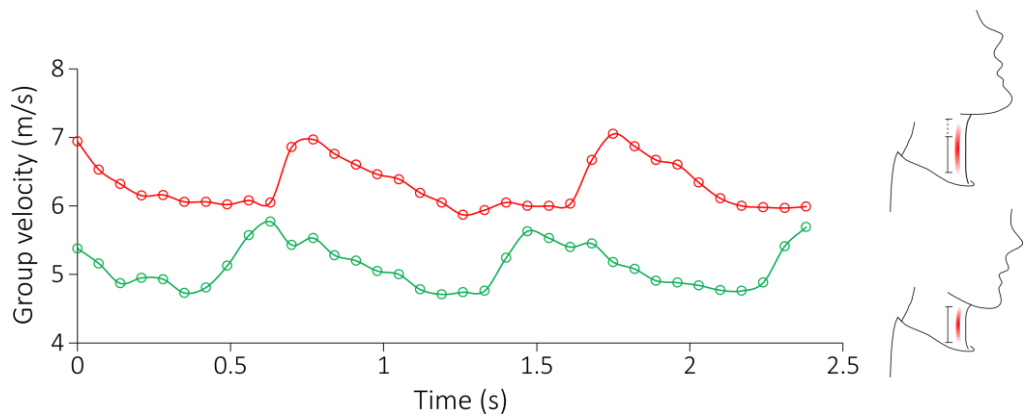

**Fig. 5 Effect of the body position on the group velocity measurements.**

The measurement of the group velocity will be affected by the body position because mechanical stress can be introduced. Fig. 5 shows the two independent measurements on the same person with different body positions, from which we find that craning the neck shifts the curve dramatically. To ensure repeatability of the experiment, the body position should be carefully controlled in measurements.

Because the guided waves excited by ARF is in the low frequency regime (less than 1.6 kHz), the arterial stiffness reported here is the average shear modulus of all layers of the arterial wall. Arteries consist of three layers: the intima, media, and adventitia, each with a distinct three-dimensional organization of elastin and collagen fibrils (Humphrey et al., 2014). Preferential fiber alignment results in anisotropic mechanical properties, therefore caution must be taken when comparing the present measurements with those obtained from classical mechanical characterization methods (Fung, 1967; Hayashi et al., 1980). While the latter primarily reflects the mechanical properties of the in-plane collagen fibrils, the average shear modulus reported in this study is dominated by the interlamellar collagen and matrix, which plays a critical role in the delamination fracture of the arterial wall, i.e., aortic dissection (Sommer et al., 2008; Yu et al., 2020). Therefore, arterial stiffness probed by guided axial waves may prove useful for the prediction of the aortic dissection events.

Different from those load-bearing tissues that are subject to stable stresses and thus undergo remodeling or plastic deformation to adapt to the stress, arterial walls are bearing the periodic stress that leads to dynamic changes in stiffness. Relying on an acoustoelastic model (Li et al., 2017b) for arteries, our method is also promising to interrogate the nonlinear elasticity of the arterial wall *in vivo* by continuously exciting and measuring the guided waves in a cardiac cycle. It has been realized that the mechanical properties of the extracellular matrix are fundamental to cellular and tissue health. Characterization of the nonlinear elasticity and thus the stiffness variations of the arterial wall induced by BP thus can provide valuable information about the mechanical homeostasis *in vivo* (Humphrey et al., 2014).

## 5   Conclusions

In this paper, we propose a novel method relying on the guided elastic waves continuously excited and detected by ultrasound to probe the BP and mechanical properties of the CCA simultaneously. In a pilot study of 17 healthy volunteers, a ~20% variation in the group velocity in a cardiac cycle is observed. A linear relation between the square of the group velocity and the BP is revealed by the experimental data and FEA, which suggests the variation of the group velocity can be interpreted as the BP waveform. We further propose to use the wavelet analysis to extract the dispersion relation of the guided waves, which provides a quantitative measurement of the arterial stiffness and its variation in response to the BP variation. The analysis and methods reported here enable the measurement of local BP and elastic properties of arteries *in vivo* and therefore can provide valuable information in early diagnosis of the CVDs.

**Declaration of Competing Interest**

The authors declare no conflict of interest.

**Acknowledgements**

We acknowledge the support from the National Natural Science Foundation of China (Grant Nos. 11572179, 11172155, 11732008, and 81561168023).

**Author contributions**

G.Y.L. and Y.C. designed the study. G.Y.L., Y.J., W.X., Y.Z., and Z.Z. carried out the experiments. G.Y.L., Y.J. analyzed the data. G.Y.L. and Y.Z. performed the numerical simulations. G.Y.L. and Y.C. wrote and revised the manuscript with inputs from all co-authors.

**Appendix**

Supplementary information is available for this paper.

**Supplementary materials**

Guo-Yang Li [*, †], Yuxuan Jiang, Yang Zheng, Weiqiang Xu, Zhaoyi Zhang, Yanping Cao[†]

Institute of Biomechanics and Medical Engineering, AML, Department of Engineering Mechanics, Tsinghua University, Beijing 100084, PR China

[*]Present address: Harvard Medical School and Wellman Center for Photomedicine, Massachusetts General Hospital, Boston, MA 02117, USA

[†]Corresponding author: G.Y. Li (lgy14@tsinghua.org.cn); Y. Cao (caoyanping@tsinghua.edu.cn)

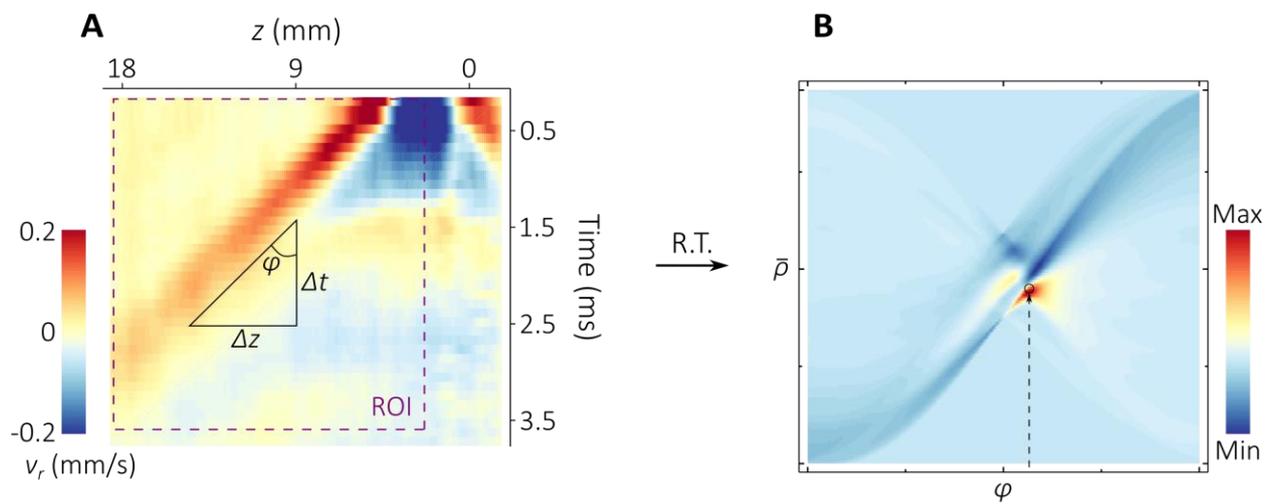

**Fig. S1 Measurement of the group velocity by Radon transformation.** (**A**) Typical time-space domain data. Only the wave propagating along positive direction of the $z$-axis is adopted in the analysis, i.e., the data in the region of interest (ROI). (**B**) Radon transformation of the data. By identifying the maximum.

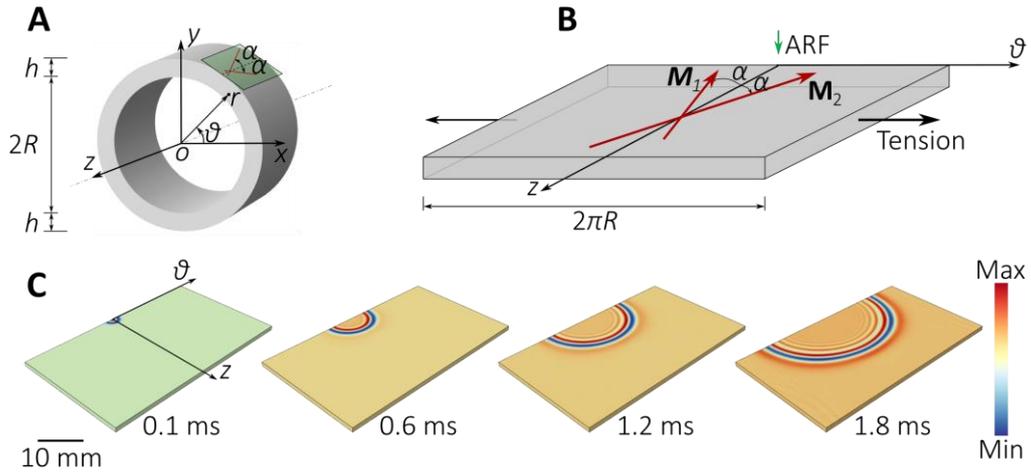

**Fig. S2 Finite element model.** (**A**) Geometrical shape of the CCA. The directions of the collagen fibers are defined within the tangent plane. $\alpha = 40.02°$. (**B**) Simplified geometry that was used in the FEA. $\mathbf{M}_1$ and $\mathbf{M}_2$ denote the directions of the fibers. Tensile stress is applied along circumferential direction. The guided axial wave was excited by the simulated ARF. (**C**) Snapshots of the wave propagation when $\sigma_{\theta\theta}/2C_{10} = 1$.

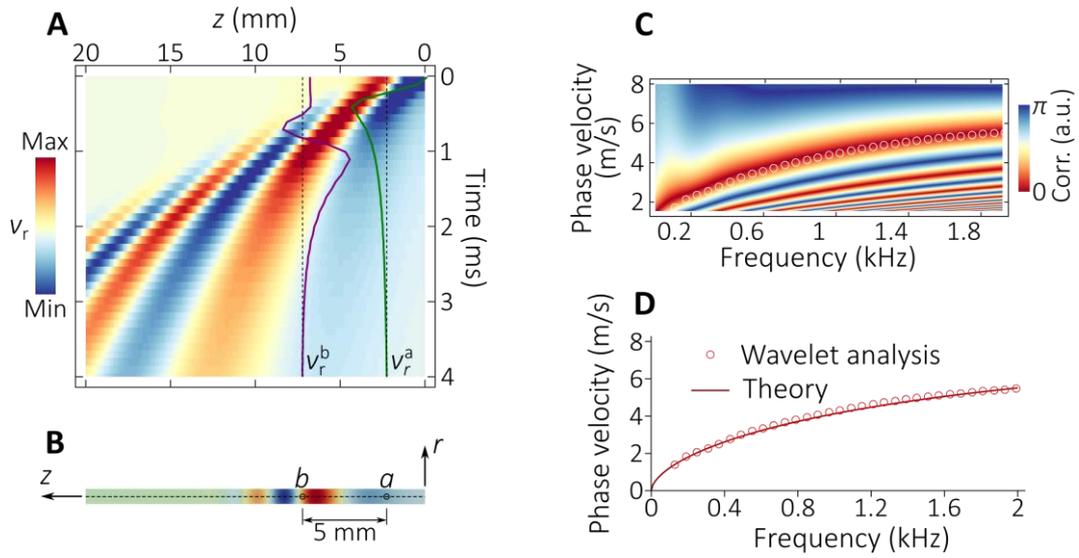

**Fig. S3 Validation of the wavelet analysis by FEA.** (**A**) Time-space domain data obtained from the FEA ($\sigma_{\vartheta\vartheta} = 0$). The two solid lines show the time profiles of $v_r$ at two different positions $a$ and $b$. (**B**) A snapshot (~0.9 ms) showing the wave propagation in the wall. The positions $a$ and $b$ are shown in the figure. The distance is 5 mm. (**C**) Cross-correlogram of $V_r^a$ and $V_r^b$ that are the wavelet transformation of $v_r^a(t)$ and $v_r^b(t)$. The white circles show the minimums in the map that correspond to the $A_0$ mode. The dispersion relation of the $A_0$ mode thus is extracted. (**D**) Comparison between the theoretical dispersion relation and that obtained from wavelet analysis. The good agreement validates our algorithm based on wavelet analysis.

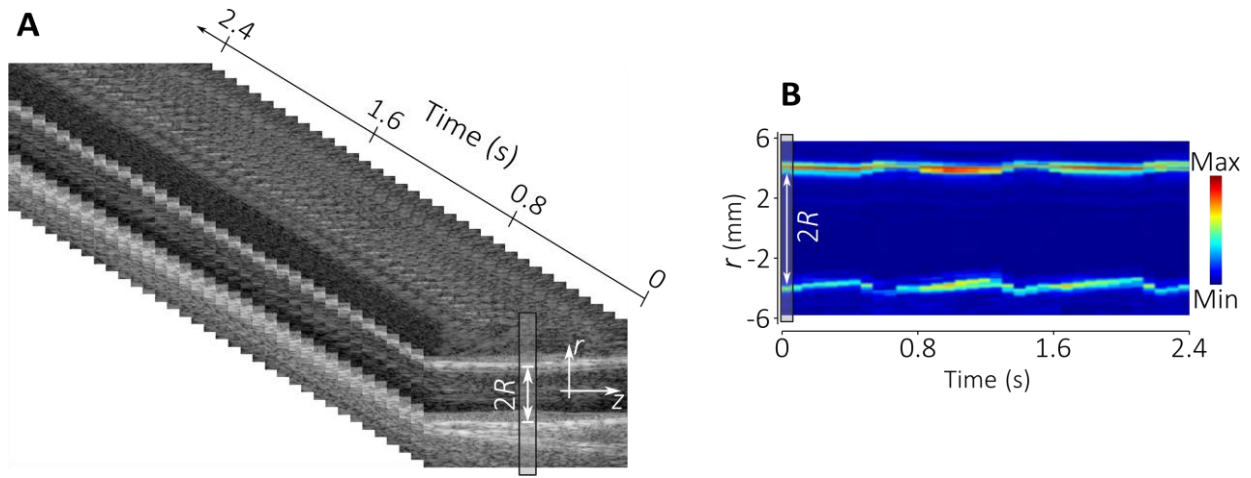

**Fig. S4 Measurement of the arterial diameter.** (**A**) Gray scale images showing the CCA. Each image denotes the first frame of each measurements. The arterial diameter is measured at the center of each frame. (**B**) The positions of the anterior and posterior walls at different times. The deformations of both the two walls are apparent when pulse wave arrives. The variation of the arterial diameter is measured from this figure.

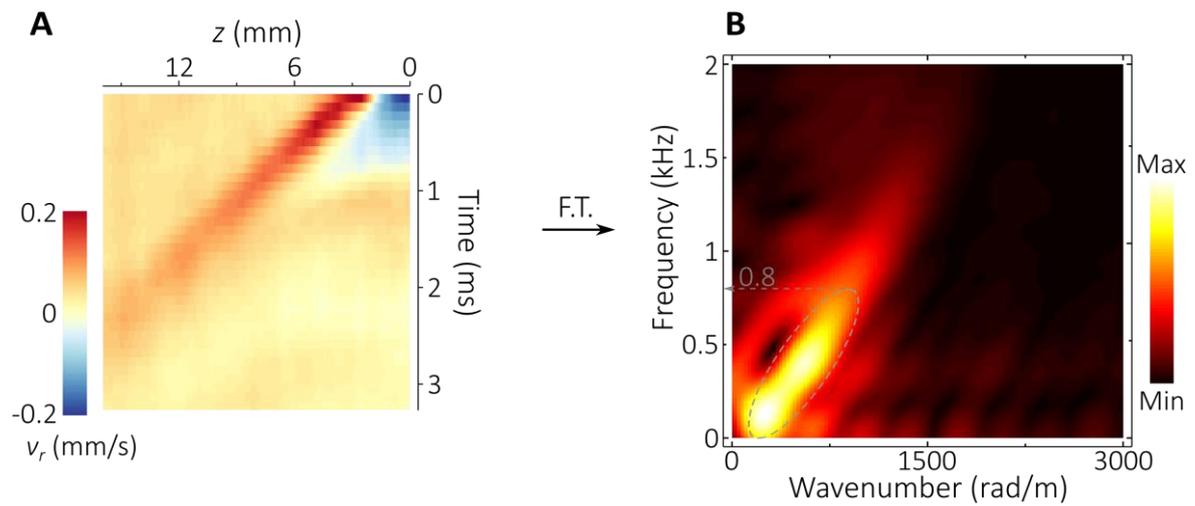

**Fig. S5 Fourier transformation of the spatiotemporal data.** (**A**) Typical time-space domain data obtained in diastole. (**B**) Two-dimensional Fourier transformation (FT) of the data. The energy is dominated by the wave modes with frequency lower than 0.8 kHz.

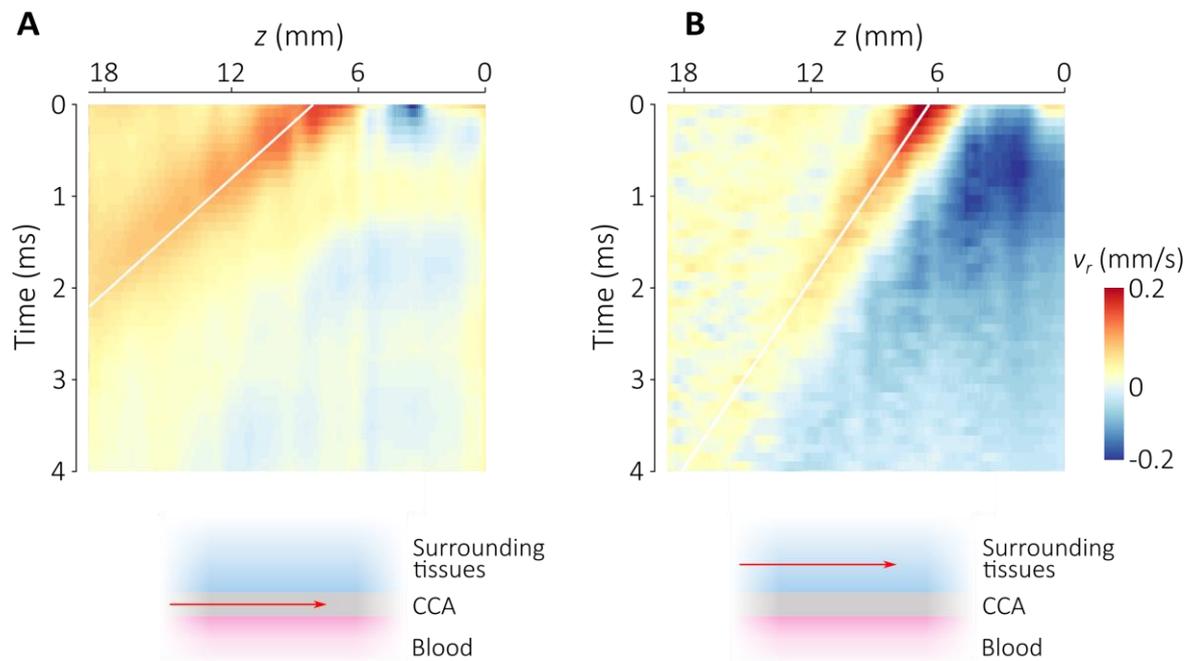

**Fig. S6 Mechanical property of the surrounding tissue.** The moving ARF was applied to excite shear waves in CCA and the surrounding tissue simultaneously. (**A**) and (**B**) showing the time-space domain data measured in CCA and the surrounding tissue. The group velocities measured from (A) and (B) are 4.9 m/s and 3.1 m/s, respectively. The shear modulus of the surrounding tissue $\mu_s$ is thus estimated via $\mu_s = \rho v_s^2 \approx 10$ kPa, where $v_s = 3.1$ m/s and $\rho = 1000$ kg/m³, by assuming an infinite geometry of the surrounding tissue. This typical data was obtained from one of the male volunteers.

**Supplementary Note1. Dispersion relation of the three-layer model**

For the three-layer model used to fit the dispersion relation, an empirical formula in an explicit form has been obtained in Ref. (Li et al., 2019):

$$c(f; \mu_{CCA}, \mu_s) = c_0 \left[ g_1 \left(1 - e^{-\frac{hf}{\tau_1 c_0}}\right) + g_2 \left(1 - e^{-\frac{hf}{\tau_2 c_0}}\right) \right], \quad (S3)$$

where $c_0 = \sqrt{\mu_{CCA}/\rho}$,

$$g_1 \left(\frac{\mu_s}{\mu_{CCA}}\right) = 0.6333 \frac{\mu_s}{\mu_{CCA}} + 0.1735, \quad (S4)$$

$$g_2 \left(\frac{\mu_s}{\mu_{CCA}}\right) = -0.4510 \frac{\mu_s}{\mu_{CCA}} + 0.6087, \quad (S5)$$

$$\tau_1 \left(\frac{\mu_s}{\mu_{CCA}}\right) = 0.0359 \frac{\mu_s}{\mu_{CCA}} + 0.0145, \quad (S6)$$

$$\tau_2 \left(\frac{\mu_s}{\mu_{CCA}}\right) = 0.1203 \frac{\mu_s}{\mu_{CCA}} + 0.1814. \quad (S7)$$

This empirical formula works well when $\mu_s/\mu_{CCA} < 0.4$.

To fit the experimental data, we use $\mu_s = 10$ kPa. Then we report the value of $\mu_{CCA}$ when the root-mean-square error (RMSE) reaches the minimum:

$$RMSE = \sqrt{\sum_{n=0}^{20} [c(f_n; \mu_{CCA}, \mu_s) - c_{DBP}(f_n)]^2}, \quad (S3)$$

where $f_n = 0.6 + 0.05n$ (kHz), $c_{DBP}(f_n)$ denotes the dispersion relation in diastole obtained from experiment.